\documentclass[twoside,leqno,twocolumn]{article}
\usepackage{ltexpprt}
\usepackage{graphicx}
\usepackage{pgfplots}
\pgfplotsset{compat=1.12}
\usepackage{filecontents}
\usepackage{multirow}
\usepackage{url}

\makeatletter
\def\footnoterule{\kern-.4\p@
    \hrule\@width 5pc\kern3\p@\kern-\footnotesep}
\begin{document}
\makeatother

\title{\Large Dynamic Analysis of Executables to Detect and Characterize Malware
	\thanks{Supported by Sandia National Laboratories' Laboratory Directed Research and Development Program,
		Hardware Acceleration of Adaptive Neural
		Algorithms (HAANA) Grand Challenge Project. 		
		Sandia National
		Laboratories is a multimission laboratory managed and operated by
		National Technology and Engineering Solutions of Sandia, LLC., a
		wholly owned subsidiary of Honeywell International, Inc., for the
		U.S. Department of Energy's National Nuclear Security Administration
		under contract DE-NA0003525. This paper describes objective technical results and analysis. Any subjective views or opinions that might be expressed in the paper do not necessarily represent the views of the U.S. Department of Energy or the United States Government. SAND NO. 2018-11011 C}}
\author{Michael R. Smith\thanks{Sandia National Laboratories. e-mail:msmith4@sandia.gov.} \\
\and
Joe B. Ingram\thanks{Sandia National Laboratories. e-mail:jbingra@sandia.gov.} \\
\and
Christopher C. Lamb\thanks{Sandia National Laboratories. e-mail:cclamb@sandia.gov.} \\
\and
Timothy J. Draelos\thanks{Sandia National Laboratories. e-mail:tjdrael@sandia.gov.} \\
\and
Justin E. Doak\thanks{Sandia National Laboratories. e-mail:jedoak@sandia.gov.} \\
\and
James B. Aimone\thanks{Sandia National Laboratories. e-mail:jbaimon@sandia.gov.} \\
\and
Conrad D. James\thanks{Sandia National Laboratories. e-mail:cdjame@sandia.gov.}}
\date{}

\maketitle







\begin{abstract} \small\baselineskip=9pt
Malware detection and remediation is an on-going task for computer security and IT professionals.
It is needed to ensure the integrity of systems that process sensitive information and control many aspects of everyday life.
We examine the use of machine learning algorithms to detect malware using the system calls generated by executables---alleviating attempts at obfuscation as the behavior is monitored rather than the bytes of an executable.
We examine several machine learning techniques for detecting malware including
random forests,
deep learning techniques, and
liquid state machines.
The experiments examine the effects of concept drift on each algorithm to understand how well the algorithms generalize to novel malware samples by testing them on data that was collected \textit{after} the training data.
The results suggest that each of the examined machine learning algorithms is a viable solution to detect malware---achieving between 90\% and 95\% class-averaged accuracy (CAA).
In real-world scenarios, the performance evaluation on an operational network may not match the performance achieved in training.
Namely, the CAA may be about the same, but the values for precision and recall over the malware can change significantly.
We structure experiments to highlight these caveats and offer insights into expected performance in operational environments. 
In addition, we use the induced models to gain a better understanding about what differentiates the malware samples from the goodware, which can further be used as a forensics tool to understand what the malware (or goodware) was doing to provide directions for investigation and remediation.

\end{abstract}

\section{Introduction}
\label{sec:intro}
Identifying malicious software (malware) on a host machine is a critical task in maintaining a system's integrity and the integrity of the work performed on that system.
Intrusion detection systems (IDSs)---such as anti-virus software---are used to identify, assess, and report any unauthorized programs on a system.
Malware authors use various techniques to evade detection by an IDS such as changing which registers are used, changing machine instructions to equivalent ones, reordering independent instruction blocks, and inserting no-operation instructions \cite{You2010_BWCCA}.
Signature-based approaches used by many IDSs are static and unable to adapt to the dynamic approaches implemented in malware, other than by repeatedly adding new signatures manually. 
As the number of computing devices increases, especially those used to process sensitive tasks (e.g., banking, health care, and infrastructure), it is imperative to detect compromised systems as soon as possible.

Despite exploiting different vulnerabilities and employing obfuscation techniques, most malware exhibits common behavior.
For example, once a system is exploited, malware will often beacon out to a command and control server or clean up log files to cover its tracks or another such activity as depicted in the cyber kill chain for advanced persistent threats \cite{hutchins2011_IWSR}.
In addition to the behavioral extent of malware, it is also important to consider the implications of concept drift wherein a target distribution (in this case, the behavior of malware) is non-stationary and changes over time.

\begin{figure}
	\centering
	\includegraphics[width=3.25in]{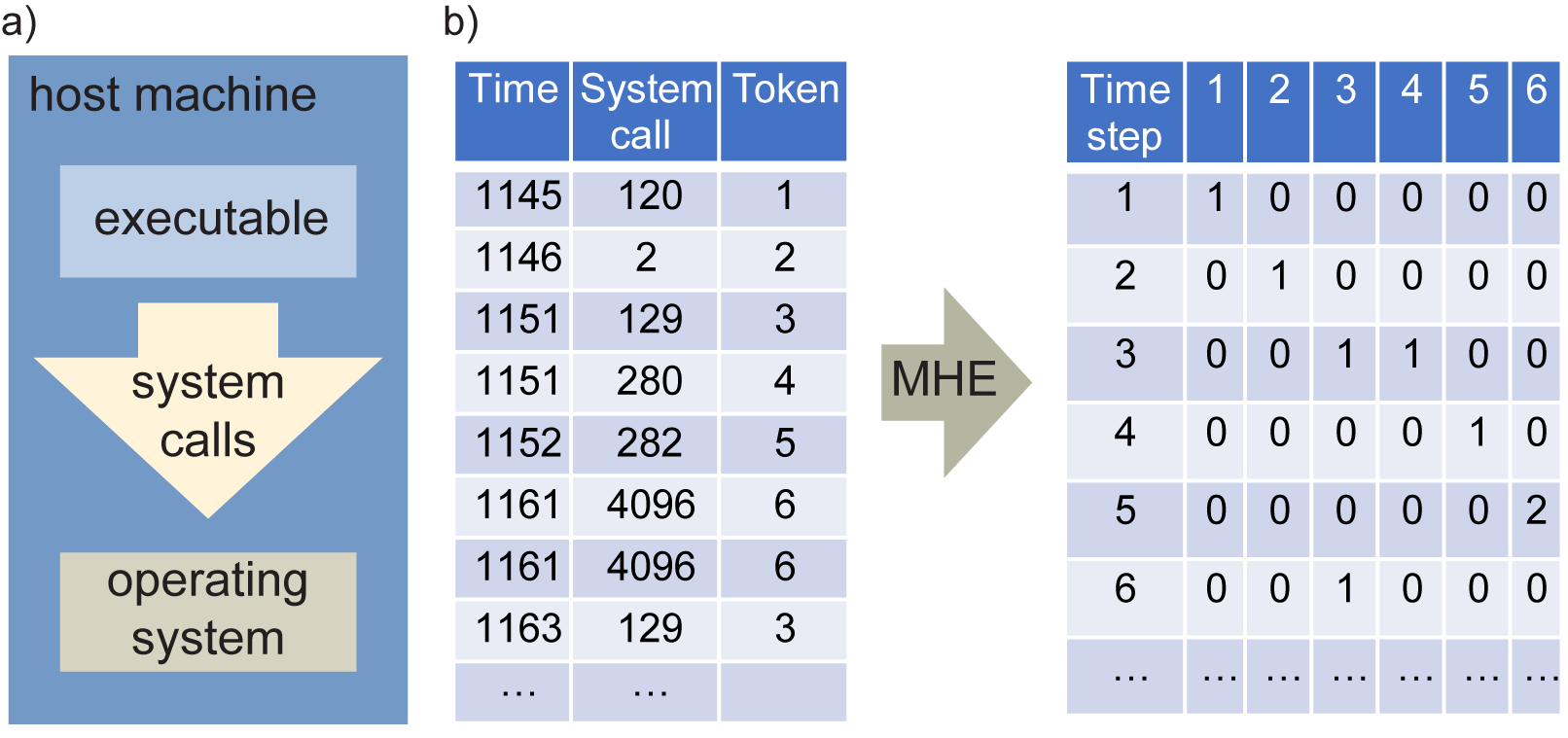}
	\caption{Process for generating the system call data. a) System calls are functions called from an executable to the underlying operating system. b) The system calls are intercepted and saved in a file and then converted to a multi-hot encoding (MHE) of the system call traces.}
	\label{fig:syscallData}	
\end{figure}

We examine machine learning (ML) algorithms (random forests (RFs), deep learning approaches and liquid state machines (LSMs)) to detect malicious behavior using system call traces (calls to functions provided by the operating system---see Figure \ref{fig:syscallData}).
We examine these methods using a concept drift scenario, whereby training data precedes test data collected from a corporate gateway.
We observe that these algorithms are able to distinguish between malware and goodware with an average class-averaged accuracy (CAA) of 93\% with a malware precision of 93\% and 88\% recall.

ML for identifying malware has several unique challenges.
Explicitly, malware authors actively try to masquerade their malware as goodware.
There is also a scaling issue in the sheer amount of data samples, the size of each data sample, and class imbalance. 
In operational networks of large corporations, there will be large amounts of executables observed across the network with less than 1\% of them being malware.
We structure experiments to investigate these issues and offer insights from traditional training schemes.

An induced model captures the generalizations of the underlying data that it is modeling.
We explore the use of ML explainability techniques to understand the characteristics of malware and identify behaviors that lead to its classification as malware.

Our contributions include:
1) a comparison of several ML models in malware identification using system call traces on real-world data,
2) an analysis of malware including which features are the most important for identifying malware, and 
3) practical insights in how to apply these results in real-world scenarios and the implications of using each technique.

\section{Related Work}
\label{sec:related}

Several previous studies have shown promising results using ML algorithms to detect malware \cite{Ranveer2015_IJCA} or to differentiate between different families of malware \cite{kolosnjaji2016_AJCAI}.
There are two common approaches to extract features from malware: static analysis and dynamic analysis \cite{Egele2008_AMCCS}.
Static analysis refers to extracting statistics from the meta-information of an executable \textit{without} running the executable, such as a list of DLLs in the binary \cite{Schultz2001_IEEESSP} or 
byte \textit{n}-grams \cite{Kolter2006_JMLR}. 
Features from static analysis are vulnerable to obfuscation techniques such as code transformation techniques, but have the advantage that malware never has to be executed.
Feeding these features into various ML algorithms produced good results showing the benefits of using ML to detect malware.
Despite this success, Kruegel et al. showed that advanced, semantic-based malware detectors can be evaded using obfuscation techniques commonly employed by malware authors.
They concluded that static analysis techniques alone are not sufficient to identify malware \cite{Moser2007_ACSAC}.

Dynamic analysis, on the other hand, runs an executable to extract features. 
Dynamic analysis, in principle, should be less vulnerable to obfuscation as it extracts features from the behavior of an executable.
Several previous works used Markov chains to model system call sequences.
Ravi and Manoharan use third-order Markov chains to model the system call traces and achieve better detection rates than support vector machines, decision trees, and na\"{i}ve Bayes \cite{Ravi2012_IJCA}.
Anderson et al. extract Markov chains of the instruction traces as features. 
Graph kernels are then used to create a similarity matrix which is then passed to a support vector machine for classification \cite{Anderson2011_JCV}.

The success of neural networks in other application areas has motivated the use of neural methods to classify malware.
For example, Nataraj et al. treat executables as gray-scale images and use well proven image processing techniques to classify malware \cite{Nataraj2011_ISVCS}.
Tobiyama et al. use long short-term memory neurons to learn a language model of the malware based on their system calls using unsupervised learning \cite{Tobiyama2016_COMPSAC}.
The output from this learned language model is then fed into a convolutional neural network for classification.
These works report accuracies of 96-98\%; however they only experimented using cross-validation, discarding any temporal ordering and not validating how the models handle concept drift.
Our work builds on the success of the previous works, provides a comparison of several methods, and examines how the investigated methods handle concept drift.

\section{System Call Data}
\label{sec:data}
System calls are the standardized programmatic pathways that allow programs to interact with the operating system.
Programs use system calls to request and manipulate computer resources controlled by the operating system, such as files, memory and network connections.
The data used for this analysis is a sequence of system calls made to the operating system from an given executable
Figure~\ref{fig:syscallData} a shows this process.
The executables were gathered from two sources.
The benign executables were pulled from the gateway of a corporate network under the assumption that the majority of the downloaded executables are benign.
These executables were run through several
anti-virus programs and, if none hit, then the sample was considered benign.
Of course, it is possible that there are malware samples in the collected executables with some probability.
This is consistent with real world situations where labels are not precise nor available for all samples.
The malware samples were gathered from daily feeds from Arbor Networks, a cybersecurity company that maintains a repository of malware. 

The data was collected over the course of 2012.
All samples are Windows 32-bit executables.
Each sample was executed in a
hypervisor environment (a platform for running virtual machines) for a given period of time and the system calls made by the top-level process were collected.
In total, 14,483 samples were collected from 6,197 benign executables and 8,286 malware samples.
In our analysis, we only analyzed the original, primary process, ignoring any spawned processes.
Future work will include an analysis of the spawned processes in addition to the parameters that are passed to the system calls.

Generally, complexity for the sequence learning methods increases with the length of the system call trace.
We limit our analysis to look at the first $n$ system calls made by an executable.
Most analyses are done on the first 1000 system calls.
An analysis of the length of the system call traces is provided in Section \ref{sec:lengthresults}.

As multiple system calls can occur in a time step (the granularity of each time step is one millisecond), we use a multi-hot encoding scheme as depicted in Figure~\ref{fig:syscallData}b.
In multi-hot encoding, a vector the length of the number of unique system calls is initialized with all zeros.
The number of times each system call made during a time step is put into the vector at the index representing the given system call.
With this base data set, we allow each algorithm to further process the data to highlight its strengths.





\section{Learning Algorithms}
\label{sec:algorithms}



\subsection{Histograms and Random Forests (Hist+RF)}
\label{sec:RF}
One way to encode the time series data for learning is to use histograms.
Each feature represents a particular system call and the value corresponds to the number of times that call was made for a given system call trace.
With this encoding, any supervised ML algorithm can be applied to the resulting dataset.
We use this encoding as input to a random forest (RF) \cite{Breiman2001_ML}.

This encoding does not capture the temporal information---the order in which system calls were made---of the data.
It can be modified to encode some ordering by using of $n$-grams, where each feature becomes a unique sequence of $n$ system calls present in the data.
However, the $n$-gram encoding increases both the dimensionality and the sparseness of the data as $n$ is increased, which can make learning more difficult.
We do not examine the use of $n$-grams here. 

\subsection{Deep Learning Methods}
\label{sec:deeplearning}
Deep neural networks have been shown to be effective for pattern recognition in images, speech, and text.
We examine the use of convolutional neural networks (CNNs) and long short-term memory (LSTM) recurrent neural networks (RNNs).
CNNs are essential elements of state-of-the-art systems for classifying objects in images and LSTMs are used in state-of-the-art text/natural language processing applications.
CNNs and LSTMs are used together in state-of-the-art speech recognition systems to capture the sequence of spectral patterns over time from speech data \cite{Amodei2016_PMLR}.
We examine the effectiveness of CNNs, LSTMs and a combination of the two for detecting malware from system call traces.


\subsubsection{Convolutional Neural Networks (CNN)}
A convolutional layer in a deep neural network learns patterns of local structure in the input signal.
Subsequent convolutional layers learn combinations of features detected in previous layers. 
Thus, the CNNs can learn feature representations over a sequence of input data \cite{Lecun1998_IEEE}.
The final layer of the CNN classifies an input sequence as goodware or malware as a function of the high-level system call structure detected over the entire sequence.
The system calls are translated into integer values in a one-dimensional vector and are then processed by a CNN with one-dimensional convolutional layers.


We tested three different architectures with a variety of kernel sizes. 
Each architecture ran for 60 epochs with kernels of five, seven, and ten elements.
All networks used categorical cross entropy as the loss function and used 32, 64, or 128 filters.
We used 
1) a CNN with two, one-dimensional convolutional layers followed by pooling, dropout, and dense layers,
2) a pure CNN with five convolutional layers, and 3) a hybrid CNN with five convolutional layers separated by batch normalization layers. 
From empirical analyses, the hybrid convolutional network had the best performance and fastest training.
Only the results from the hybrid model are reported.

\subsubsection{Long Short-Term Memory (LSTM)}
For the LSTM \cite{Hochreiter1997_NerualComp}, each system call trace includes a time-ordered sequence of system call IDs on which an LSTM network can detect temporal (sequence) patterns that are important for discriminating between goodware and malware.
One and two layers of LSTM neurons with various numbers of neurons per layer were explored. Each node learns a different sequence pattern and the collection of sequence pattern detectors from all the nodes connected to the output layer are used to classify each system call log.

\subsubsection{Combined Convolutional and Recurrent Neural Network (CNN+LSTM)}
Combining layers of a CNN with LSTM layer(s) has been shown to be a powerful classifier that learns temporal patterns in sequences of local structure. Convolutional layers can present a sequence of higher-level features to an LSTM layer, which often leads to superior performance than an LSTM presented with raw sequence data.
Thus, we examine using a convolutional layer before feeding the input into 1 or 2 LSTM layers.

\subsection{Liquid State Machines (LSM)}
The LSM \cite{Maass2002_NeuralComput} is a neural-inspired algorithm that mimics the cortical columns in the brain.
LSMs are composed of three general components:
1) input neurons,
2) randomly connected leaky integrate-and-fire spiking neurons (LIF) called the liquid, and
3) readout nodes that read the state of liquid.
The liquid functions as a temporal kernel, casting the input data into a higher dimension and the LIF neurons allow for temporal state to be carried from one time step to another.
We use a liquid of 135 neurons where the inputs are randomly connected to 30\% of the neurons in the liquid.



The readout neurons are the only neurons that have plastic synapses, allowing for synaptic weight updates via training.
Any classifier can be used, but often a linear classifier is sufficient.
We use a support vector machine with a radial basis function kernel to train the readout neurons.
The sigma and box parameters for the kernel are chosen using Bayesian optimization minimizing the 10-fold cross validation loss.


One benefit of using a liquid state machine is that it can be run on neuromorphic hardware, which will significantly reduce the computational time and power consumption \cite{Smith2017_IJCNN}.

\section{Experimental Methodology}
\label{sec:methods}

To evaluate each algorithm described in Section \ref{sec:algorithms}, we report the accuracy (Acc), the class averaged accuracy (CAA), and, for the malware class, the precision (MPr) and recall (MRe).
The CAA is the average of the accuracy for each class.
We split the data into training and test sets maintaining temporal ordering (the instances in the training set were observed before the instances in the test set).
The temporal ordering allows for a test of how well the models handle concept drift.
The distributions for the goodware and malware are shown in the sorted column in Table \ref{tab:data}.
The distributions in the distributed column are used in a later experiment mimicking operational network traffic characteristics.

\begin{table}
	\begin{center}
		\textbf{Number of Samples in Datasets}
		\begin{tabular}{l|ccccc}
			& \multicolumn{2}{c}{Goodware} \\
			& Distributed & Sorted \\
			\hline
			Training 		&  11757 	& 13265 \\
			Testing 		& 	4728	& 3220 \\
			\hline
			& \multicolumn{2}{c}{Malware} \\
			\hline
			Training 		& 11091 	& 9092 \\
			Testing 		& 45 		& 2044 \\
		\end{tabular}
	\end{center}
	\caption{The data distributions used in the distributed (down-sampled) and sorted data sets. Both of these sets preserve temporal ordering between the training and test sets.}\label{tab:data}
\end{table} 

In current IDSs, the number of false positives can and often does overwhelm an analyst.
The analyst often has to manually investigate any alerts from the IDS to understand if a compromise has occured.
Thus, reducing the number of false positives by an order of magnitude significantly reduces the work load for an analyst.
Also, intrusion detection requires high recall as only one vulnerability needs to be exploited for an adversary to accomplish his or her objective. 
Thus, the requirements for malware detection reach far beyond classification accuracy.


\section{Experimental Results}
\label{sec:results}

The results for examining the first 1000 systems calls are shown in Table \ref{tab:algResults} with the highest performing methods highlighted in bold for each metric.
Each method is able to distinguish between malware and goodware with 90\% CAA or greater.
Overall, the Hist+RF achieves the highest measures and is comparable to using a voting ensemble of all the methods (including Hist+RF).
This result is surprising as we anticipated that the sequence learning methods would outperform a histogram representation of the data.
We provide a further analysis of this result in Section \ref{sec:featureRep}.

\begin{table}
	\begin{center}
		\begin{tabular}{l|cccc}
			Alg & Acc & CAA & MPr & MRe\\
			\hline
			Hist+RF & \textbf{95.3} & 94.7 & 0.953 & \textbf{0.926} \\
			CNN & 94.0 & 93.2 & 0.946 & 0.896 \\
			LSTM & 91.3 & 90.0 & 0.926 & 0.843 \\
			CNN+LSTM & 94.5 & 93.7 & 0.956 & 0.901 \\
			LSM & 90.7 & 89.8 & 0.856 & 0.856 \\
			\hline
			Ensemble & \textbf{95.3} & \textbf{95.5} & \textbf{0.962} & 0.917 \\
		\end{tabular}
		\caption{The accuracy (ACC), class averaged accuracy (CAA), malware precision (MPr) and malware recall (MRe) on sequence lengths of 1000. The highest value for each metric is bolded.}
		\label{tab:algResults}
	\end{center}
\end{table}

The results are statistically significant using Cochran's Q test---a non-parametric test for measuring the differences between three or more measurements.
(The null hypothesis is that there is no difference between classifier outputs.)
With the null hypothesis rejected, we test the significance between pairs of algorithms using the pairwise Cochran's Q test---equivalent to McNemar's test---and use the \v{S}id\'{a}k correction to control for the family-wise error rate.

Table \ref{tab:significance} shows which pairs of algorithms are different with statistical significance.
With an alpha value of 0.05, the ensemble is statistically significantly different from all of the other methods.
Despite the Hist+RF achieving higher accuracy, it is only significantly different from the LSTM.
Interestingly, the LSTM is significantly different from all of the other algorithms except the LSM. 
This may be due to the neurons in both algorithms maintaining temporal state.
The CNN+LSTM is significantly different from the LSTM and LSM.
This gives some insight into the power of the input representation from the convolutional layer in CNNs and CNN+LSTMs.

\begin{table}
	\begin{center}
		\begin{tabular}{l|ccccc}
			& Hist+ & & & CNN+ & \\
			& RF & CNN & LSTM & LSTM & LSM \\
			\hline
			Ensemble & \textbf{YES} & \textbf{YES} & \textbf{YES} & \textbf{YES} & \textbf{YES} \\
			Hist+RF & - & NO & \textbf{YES} & NO & NO \\
			CNN & NO & - & \textbf{YES} & NO & NO \\
			LSTM & \textbf{YES} & \textbf{YES} & - & \textbf{YES} & NO \\
			CNN+LSTM & NO & NO & \textbf{YES} & - & \textbf{YES} \\
			LSM & NO & NO & NO & \textbf{YES} & - \\
		\end{tabular}
		\caption{Pair-wise comparison of the investigated algorithms for statistical significance using the McNemar test with an alpha-value of 0.05 and \v{S}id\'{a}k correction.}
		\label{tab:significance}
	\end{center}
\end{table}

The success of CNNs on system call log classification was surprising given that there is no identifiable, quantitative relationship between neighboring system call IDs as there is with image pixels, for example.
Nonetheless, repeated, local, discriminative patterns exist in the log files that the CNNs are able to learn.
The poorer performance of networks consisting only of LSTM layers may be due to the embedding used and/or inadequate training data to learn the temporal patterns necessary for good classification.
For high-performing object recognition in images, state-of-the-art models use data augmentation where there are instances of the same object with different aspects including lighting, scales, and orientations.

For sequence data, the problem can be compounded by a separation in time or the re-ordering of important sequence structures.
It is possible that some types of malware and goodware did not have a rich enough representation in the training set for the LSTMs to adequately learn to recognize them.
Data augmentation, which we did not do, is often a necessary element to creating a high-performance pattern recognizer.
Data augmentation is difficult and can be especially difficult with cyber data since exact values can be important and should not be perturbed by random noise.
An ideal situation would be to know the sequence patterns that discriminate between malware and goodware and to surround those patterns with various types of background data and separate them by random lengths of time.

Overall performance of the CNNs hint at a potentially larger feature space than the dimensions of the selected kernel for the CNN.
The best CNN performance resulted from a hybrid network with five layers, downsampling, and normalization between each layer.
This strategy provides some amount of resistance to noise while allowing the CNN to detect features over longer sequences.
This also seems to align with the performance of Hist+RFs.

The high performance of RFs was surprising.
RFs are a simpler approach for classification and are easier to train and maintain than neural techniques.
All approaches perform well, however, and could be used in a variety of domains based on specific domain needs.

\subsection{System Call Length}
\label{sec:lengthresults}

\begin{figure}
	\begin{center}
		\begin{tikzpicture}
		\begin{axis}[
		width=0.475\textwidth,
		height=6cm,
		ymajorgrids,
		x tick label style={/pgf/number format/1000 sep=},
		ytick scale label code/.code={$\times$},
		legend style={at={(0.7,0.55), anchor=south east}},
		xlabel=Seqence Length,
		ylabel=CAA
		]
		\addplot[very thick] table[x=length,y=LSM] {data.csv};\addlegendentry{LSM}
		\addplot[no marks, dashed, blue, very thick] table[x=length,y=RF] {data.csv};\addlegendentry{Hist+RF}
		\addplot[no marks, densely dotted, red, very thick] table[x=length,y=CNN] {data.csv};\addlegendentry{CNN}
		\addplot[no marks, dashdotted, green, very thick] table[x=length,y=CNN+LSTM] {data.csv};\addlegendentry{CNN+LSTM}
		\addplot[no marks, dashdotted, very thick ] table[x=length,y=LSTM] {data.csv};\addlegendentry{LSTM}
		\end{axis}
		
		\end{tikzpicture}
		\caption{The CAA over different lengths of the system call traces varying between 100 and 5000.}
		\label{fig:seqLength}
		
	\end{center}
\end{figure}
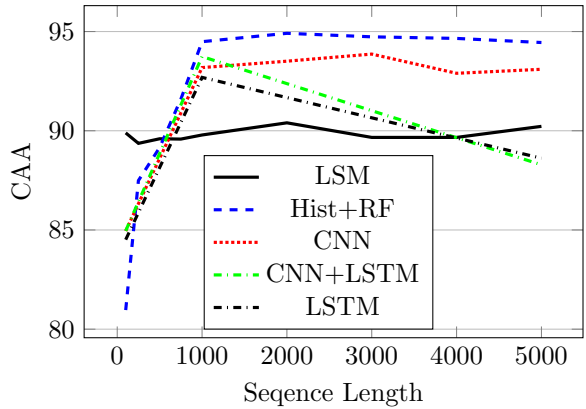
We examined how the sequence length affects the overall performance.
The CAAs for each algorithm are shown in Figure \ref{fig:seqLength}.
With only 100 system calls, the LSM has the highest CAA at 90\%.
The CAA for the LSM stays around 90\% regardless of the sequence length..
The CAAs for the other methods increase as the number of system calls increases up to a length of 1000. 
The fact that the LSM is able to achieve high CAA with only 100 system calls may be due to the signal averaging out as the number of system calls increases. 
It is also interesting that with only 100 system calls, malware can be identified with high accuracy. 

Including more than 1000 system calls does not provide a significant improvement in CAA.
Thus, we can surmise that sufficient information is provided in the first portion of an executable for differentiating between goodware and malware.

\begin{table}
	\begin{center}
		\begin{tabular}{l|l|cccc}
			Alg & Data & CAA & Acc & MPr & MRe \\
			\hline
			\multirow{3}{*}{Hist+RF} & Sort & 95.3 & 94.7 & 0.953 & 0.926 \\
			& CV & \textbf{96.3} & 96.0 & \textbf{0.965} & 0.942 \\
			& Dist & 95.9 & \textbf{97.3} & 0.187 & \textbf{1.000} \\
			\hline
			\multirow{3}{*}{CNN} & Sort & 94.0 & 93.2 & 0.946 & 0.896 \\
			& CV & 95.5 & 95.1 & \textbf{0.959} & 0.928 \\
			& Dist & \textbf{97.0} & \textbf{98.5} & 0.242 & \textbf{1.000} \\
			\hline
			\multirow{3}{*}{LSTM} & Sort & 91.3 & 90.0 & \textbf{0.926} & 0.843 \\
			& CV & 90.9 & 90.0 & 0.850 & 0.919 \\
			& Dist & \textbf{92.4} & \textbf{94.0} & 0.107 & \textbf{0.956} \\
			\hline
			\multirow{3}{*}{CNN+LSTM} & Sort & 94.5 & 93.7 & \textbf{0.956} & 0.901 \\
			& CV & 94.8 & 94.2 & 0.955 & 0.914 \\
			& Dist & \textbf{95.0} & \textbf{96.4} & 0.157 & \textbf{0.978} \\
			\hline
			\multirow{3}{*}{LSM} & Sort & 90.7 & 89.8 & 0.856 & 0.856 \\
			& CV & \textbf{93.1} & 92.6 & \textbf{0.926} & 0.901 \\
			& Dist & 91.3 & \textbf{95.6} & 0.098 & \textbf{1.000} \\
		\end{tabular}
		\caption{A comparison of results from diffenent evaluations of each algorithm: a) sorted data set with roughly balanced goodware to malware ratio, b) 10-fold cross-validation, and c) a test set with significant class skew. }
		\label{tab:comp}
	\end{center}
\end{table}

\subsection{Generalizing the Results to Real-World Scenarios}
Up to this point, the examined data set is temporally structured, but it contains a fairly balanced distribution of goodware to malware in the test set.
While this allows for an evaluation of a broad spectrum of malware samples, it is not representative of the distribution of malware found in operational networks where the ratio of malware to goodware is much lower.
To address this, we also created a data set with a skewed data set as shown in Table \ref{tab:data} in the distributed column (a random down-select).

In typical operational situations, ML algorithms are first evaluated on a training set---often balanced as done here as class skew has been shown to exacerbate the effects of other characteristics causing misclassifications \cite{Smith2014_ML}.
A common approach for evaluating approaches is to use \textit{n}-fold cross-validation.
However, this can provide an overly optimistic evaluation of the performance as temporal ordering is removed.
Deploying to a different distribution than what was used for testing can result in significantly different performance.

Table \ref{tab:comp} shows the results of evaluating the investigated algorithms using the sorted data set that we have been examining, 10-fold cross-validation, and a distributed data set with significant class skew in the test set as might be observed in an operational network.
Generally, cross-validation achieves better metrics than the sorted data set.
The cross validation results do not take concept drift into account in the performance metric as the temporal ordering is \textit{not} preserved.
This shows that concept drift is an important aspect to take into account when developing models for malware detection.
Also, if using cross-validation, the same performance levels should not be expected in operational settings.

For the distributed data set, the CAA is often similar to the cross-validation and the sorted data set but precision and recall on the malware is significantly different---recall is close to 1 and precision is very low.
The low precision is due to the fact that the ratio of malware to goodware is lower in the distributed scenario.
In our case, we have 45 malware and 4728 goodware samples.
If only 3\% of the goodware is misclassified, then 142 samples are misclassified and the malware precision is 24\%.
For the sorted scenario with more malware samples, the malware precision changes to almost 95\%.
Thus, the effect of class skew alone has dramatic affects on the results despite achieving good results from cross-validation and using the sorted data set.

\section{Discussion and Further Analysis}
\label{sec:analysis}
\subsection{Feature Representation}
\label{sec:featureRep}
Our results have shown that  ML is a viable option for identifying malware from system call traces.
We had hypothesized that the more sophisticated sequence learning methods would outperform the Hist+RFs.
To further investigate why the Hist+RF perform as well as it does, we test whether using a RF and/or representing the data with histograms has a significant affect on the results.
While the Hist+RF was not statistically significant when compared to the other neural algorithms (other than the LSTM) it has a much lower training complexity than the neural methods.

To investigate if the RF caused the high accuracy, we took the output from the neural methods before they are fed into the last layer and used them as input to a RF.
This tests whether the RF provides more discriminatory power than the linear classifier found at the last layer in a neural network.
It also provides insight into whether the deep learning methods are able to automatically extract higher-order features from the system call traces as has been shown in other domains.

\begin{table}
	\begin{center}
		\begin{tabular}{l|cc}
			Output rep & RF & Non-RF\\
			\hline
			CNN & 93.1 & 94.0 \\
			CNN+LSTM & 93.7 & 94.5 \\
			LSM & 89.5 & 90.7 \\
			
		\end{tabular}
		\caption{The CAA for a RF trained on the outputs before the output layer for neural methods. The right column gives the original CAA for the algorithm.}
		\label{tab:outputRep}
	\end{center}
\end{table}

The results for using the output from the last layer of the neural methods as input to a RF are shown in Table \ref{tab:outputRep}.
The RF column represents using the output from the neural models as the features input to a RF.
The Non-RF column refers to using the original classifier of the algorithm (linear classifier).
The results show that the CAAs decreases when a RF is used as the classifier, although only by about 1\%.

We also test whether the histograms were able to achieve better results using a linear classifier.
Using the histograms as input to a linear classifier results in a decrease of 10\% CAA from 95.3\% to  85.0\%.

While the system call sequences describe the behavior of an executable, there are several methods for achieving the same functionality in an executable.
Malware authors take full advantage of that fact to obfuscate the behavior of their code including adding spurious system calls and changing the order of the system calls.
This makes it difficult when trying to learn using the sequence of system calls.
The space of system call sequences is very large, yet the valid sequences of system calls (sequences that execute a valid process) are sparsely distributed throughout the space---especially those for doing malware.
The space is not continuous, making it difficult for gradient-descent based optimization-based methods to generalize.
Given more data and a variety, the sequence learners may have been able to produce better results.
These constraints may have played into why Hist+RF performed as well as it did.

\subsection{Characterizing Malware}
In addition to identifying, it is important to also understand what the malware is doing and why it is classified as malware.
There are several methods to explain what the model has learned from techniques such as feature importance and explainability.

The Gini importance value measures the importance of each feature from an induced RF \cite{Breiman1984_CART}.
We use the implementation provided in the Python package scikit-learn \cite{scikit-learn}.
The 20 most important features (system calls) using the RF with histograms are shown in Table \ref{tab:featureImportance}.
The Goodware and Malware columns represent if the system call was of also one of the 20 most frequently called system call relative to the other class.


The most discriminating system calls deal with file IO and virtual memory allocation.
Malware issues NtFsControlFile system calls more frequently than goodware.
This system call ``sends a control code directly to a specified file system or file system filter driver, causing the corresponding driver to perform the specified action" \cite{ZwFsControlFile}.
Thus, we can see that, in general, malware may try to be more specific with which resources it is using.
Incorporating the parameters sent with system calls would provide more details about its behavior.

\begin{table}
	\centering
	\begin{tabular}{l|cc}
		Feature & Goodware & Malware \\
		\hline
		NtAllocateVirtualMemory &  & X \\
		NtClose & X & \\
		NtCreateEvent & - & - \\
		NtCreateFile & X & \\
		NtCreateSection & - & - \\
		NtFreeVirtualMemory &  & X \\
		NtFsControlFile &  & X \\
		NtMapViewOfSection &  & X \\
		NtOpenKey & X & \\
		NtOpenSection &  & X \\
		NtQueryAttributesFile & - & - \\
		NtQueryInformationFile & X & \\
		NtQueryInformationProcess & X & \\ 
		NtQueryInformationToken & X & \\
		NtQuerySection &  & X \\
		NtQuerySystemInformation &  & X \\
		NtQueryValueKey & X & \\
		NtReadFile & X & \\
		NtRequestWaitReplyPort &  & X \\
		NtSetInformationFile & X & \\
	\end{tabular}
	\caption{The most importance features using the Gini importance values in Hist+RF. The Goodware and Malware columns represent whether a system call was more likely to be called by goodware or malware.}
	\label{tab:featureImportance}
\end{table}

\begin{figure*}
	\centering
	\begin{tabular}{cc}
		\includegraphics[width=3.35in]{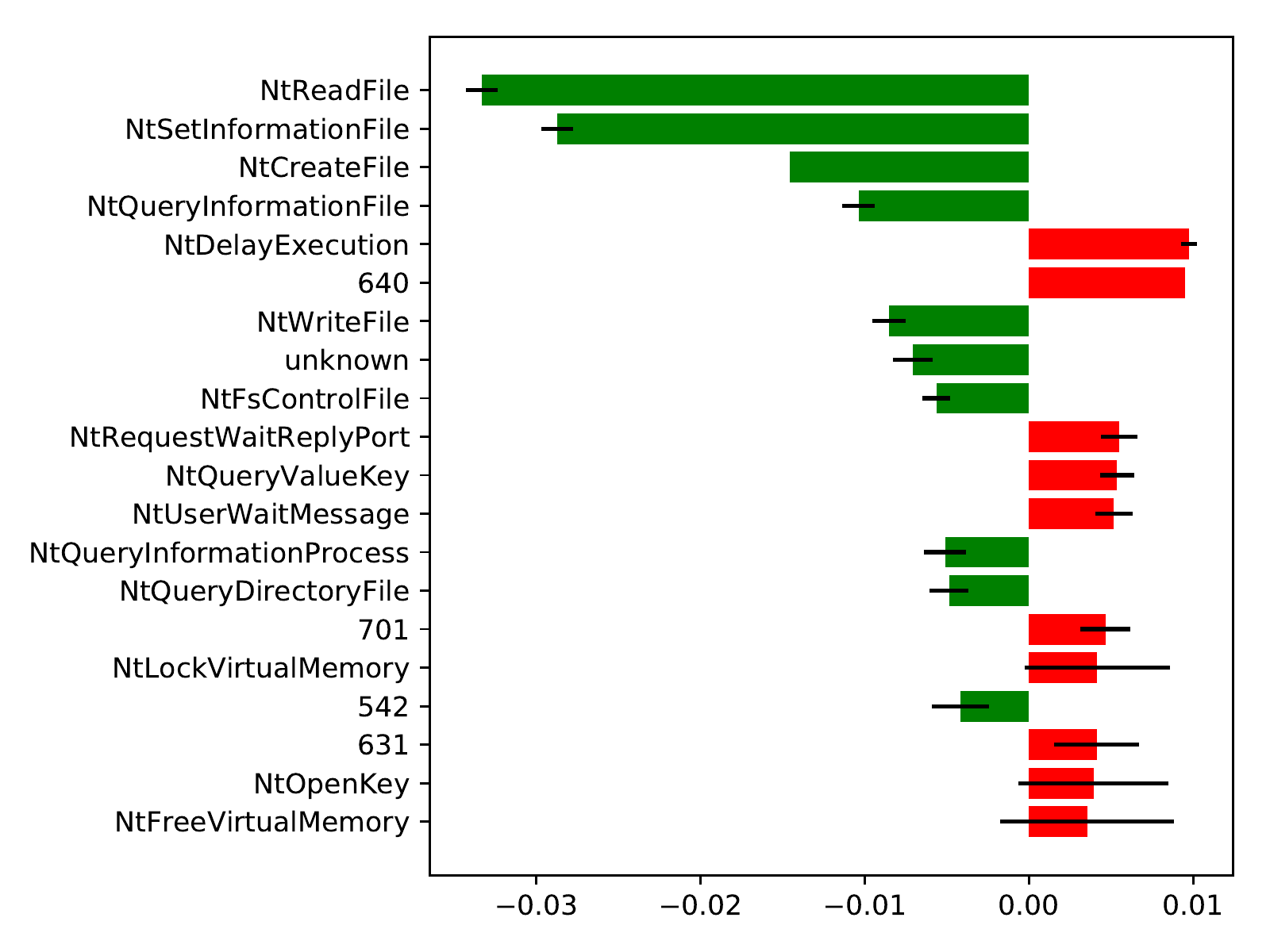} & 
		\includegraphics[width=3.35in]{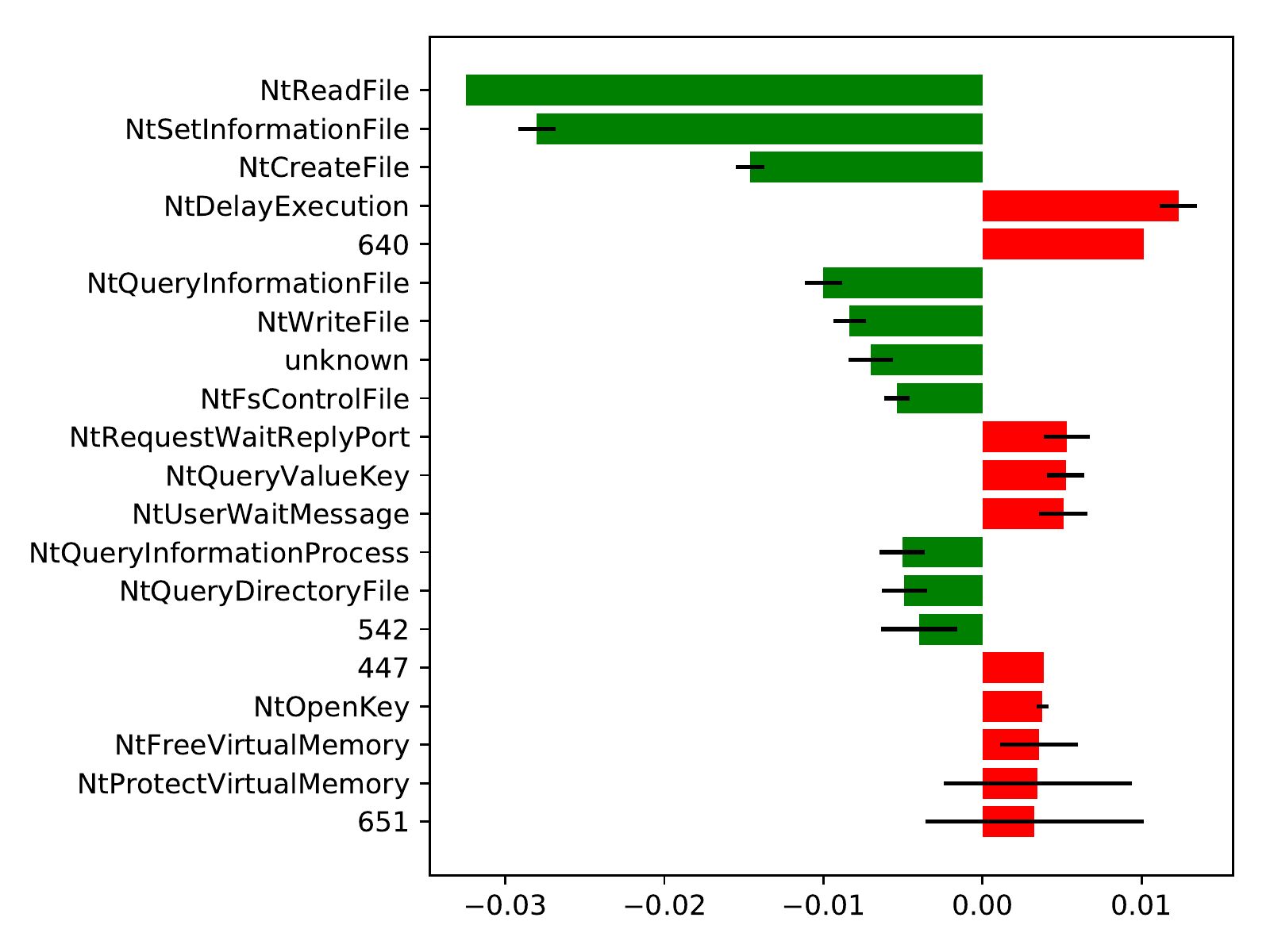} \\
		a & b \\
	\end{tabular}
	\caption{The most important features on average with standard deviation error bars using LIME for a) correctly classified malware samples and b) misclassified malware samples. The length of the bar represents how influential it is for goodware or malware. Green and to the left for malware and red and to the right for goodware.}
	\label{fig:LIME}
\end{figure*}

The Gini feature importance measure provides a global overview of the characteristics of malware and goodware.
We also examine the important features for each individual prediction using Local Interpretable Model-Agnostic Explanations (LIME) \cite{Ribeiro2016_KDD}. 
The top 15 averaged feature importance from LIME for correctly identified malware and misclassified malware are shown in Figure \ref{fig:LIME}a and \ref{fig:LIME}b, respectively.
The green bars to the left are feature importance values for malware and the red bars to the right are feature importance values for goodware.
The standard deviation error bars indicate how much variance there is when calculating the mean.
Several of the features overlap with those found by analyzing global feature importance.

The LIME explanations for the first three features are the same for the correctly classified and misclassified malware samples with little variance.
The explanations diverge starting with the fourth and fifth features, again, with little variance for correctly classified and misclassified samples.
This is important to note because the fourth and fifth features for the misclassified malware indicate that the features change the classification to goodware.
Malware authors could use this information to better obfuscate their malware while those protecting networks can use this information to improve their models.
This information could also be used for remediation purposes.
For a given sample, knowing what the malware was doing that led to it being classified as malware can provide a starting point for a forensic investigation.

Decision trees (DTs) can be used to create a set of rules to better understand the structure in the data.
We train a DT on output from the RF and examine the splits that are induced by the DT.
Rules, such as the rule shown in Figure \ref{fig:DT}, could be extracted from the DT and used to supplement other IDSs such as SNORT.

\begin{figure}
	\centering
	\small
	\begin{verbatim}
	
	if NtSetInformationFile <= 0.005,
	if NtSetInformationFile <= 0.003,
	if NtReadFile <= 0.315, 
	if NtDuplicateObject <= 0.004, 
	if NtCreateSymbolicLinkObject <= 0.001, 
	class=malware, [ 7.  2329.]
	\end{verbatim}
	\caption{Example rule from an induced decision tree.}
	\label{fig:DT}
\end{figure}

Examining the DT, the first split is on NtSetInformationFile and the majority of instances with that one split are malware (2666:176 malware to goodware).
Following the tree down further provides finer granularity.
Knowing which system calls or combinations of system calls have high discriminatory power can be very powerful for defending a system, identifying vulnerabilities, and mitigating their risk.
Using ML models with a domain expert could help to harden computer systems against malware.

\subsection{Algorithmic Considerations}
The discussion thus far has focused on the histogram represenation of the system calls.
Explaining sequences of inputs is not as well established.
In addition, there are algorithmic constraints that should be considered when deciding which algorithms to use.

The neural methods inherently face a computational bottleneck with the vector-matrix multiply.
The parameter space (number of weights) is extremely large, especially as the number of neurons increases.
In addition, there is a large hyper-parameter space (e.g. architecture of the network, activation functions, momentum, and dropout) that has a significant impact on the induced model.
Hist+RF is relatively simple algorithmically compared to the neural methods.
The neural methods also require large amounts of data to learn an effective model.
With only 10,000-13,000 training examples from a complex space and some with very few system calls (short sequences), the neural methods were not able to perform as well.
Their could possibly be improved by using unlabeled data for pre-training.

\section{Conclusion}
\label{sec:conclusion}
In this paper, we examined several ML methods as part of a dynamic analysis of executables for detecting malware.
Each proved to be a viable solution achieving over 90\% CAA.
We examined techniques for characterizing the malware using feature importance and explainability techniques.
Extending these techniques to data sequences could help better characterize malware and how to mitigate its effects on a system.
The results and techniques presented here can serve as a step to improving security against malware.

\bibliography{refs}
\bibliographystyle{siam}
\end{document}